\def\papertitle{Differentiable all-pole filters for time-varying audio systems}

\def\paperauthorA{Chin-Yun Yu}
\def\paperauthorB{Christopher Mitcheltree}
\def\paperauthorC{Alistair Carson}
\def\paperauthorD{Stefan Bilbao}
\def\paperauthorE{Joshua D. Reiss}
\def\paperauthorF{Gy{\"o}rgy Fazekas}

\documentclass[twoside,a4paper]{article}
\usepackage{etoolbox}
\usepackage{dafx_24}
\usepackage{amsmath,amssymb,amsfonts,amsthm}
\usepackage{mathtools}
\usepackage{euscript}
\usepackage[T1]{fontenc}
\usepackage[utf8]{inputenc}
\usepackage{ifpdf}
\usepackage[english]{babel}
\usepackage{caption}
\usepackage{subfig} 
\usepackage{color}
\usepackage{tabularx}
\usepackage{multirow}
\usepackage{booktabs}
\usepackage{siunitx}
\usepackage[boxed, ruled]{algorithm2e}

\input glyphtounicode
\ninept

\newcounter{numauth}
\newcounter{listcnt}
\newcommand\authcnt[1]{\ifdefined#1 \stepcounter{numauth} \fi}

\newcommand\addauth[1]{
\ifdefined#1 
\stepcounter{listcnt}
\ifnum \value{listcnt}<\value{numauth}
\appto\authorslist{, #1}
\else
\appto\authorslist{~and~#1}
\fi
\fi}
\authcnt{\paperauthorB}
\authcnt{\paperauthorC}
\authcnt{\paperauthorD}
\authcnt{\paperauthorE}
\authcnt{\paperauthorF}
\authcnt{\paperauthorG}
\authcnt{\paperauthorH}
\authcnt{\paperauthorI}
\authcnt{\paperauthorJ}
\def\authorslist{\paperauthorA}
\addauth{\paperauthorB}
\addauth{\paperauthorC}
\addauth{\paperauthorD}
\addauth{\paperauthorE}
\addauth{\paperauthorF}
\addauth{\paperauthorG}
\addauth{\paperauthorH}
\addauth{\paperauthorI}
\addauth{\paperauthorJ}

\newcommand{\q}{\mathbf{q}}

\usepackage{times}

\newif\ifpdf
\ifx\pdfoutput\relax
\else
   \ifcase\pdfoutput
      \pdffalse
   \else
      \pdftrue
\fi

\ifpdf 
  \usepackage[pdftex,
    pdftitle={\papertitle},
    pdfauthor={\authorslist},
    pdfsubject={Proceedings of the 27th International Conference on Digital Audio Effects (DAFx24)},
    colorlinks=false, 
    bookmarksnumbered, 
    pdfstartview=XYZ 
  ]{hyperref}
  \usepackage[pdftex]{graphicx}
\else 
  \usepackage[dvips]{epsfig,graphicx}
  \usepackage[dvips,
    pdftitle={\papertitle},
    pdfauthor={\authorslist},
    pdfsubject={Proceedings of the 27th International Conference on Digital Audio Effects (DAFx24)},
    colorlinks=false, 
    bookmarksnumbered, 
    pdfstartview=XYZ 
  ]{hyperref}
\fi
\usepackage[hypcap=true]{caption}
\title{\papertitle}
\usepackage{xcolor}
\def\SB[#1]{\textcolor{blue}{#1}}

\affiliation{
\paperauthorA$^\flat$\sthanks{Equal contribution.},\, \paperauthorB$^{\flat*}$,\, \paperauthorC$^\sharp$,\, \paperauthorD$^\sharp$,\, \paperauthorE$^\flat$,\, and \paperauthorF$^\flat$}
{$^\flat$\href{https://www.c4dm.eecs.qmul.ac.uk/}{Centre for Digital Music}, Queen Mary University of London, London, UK\\
$^\sharp$\href{http://www.acoustics.ed.ac.uk/}{Acoustics and Audio Group}, University of Edinburgh, Edinburgh, UK\\
{
\tt 
\{
\href{mailto:chin-yun.yu@qmul.ac.uk}{chin-yun.yu},
\href{mailto:c.mitcheltree@qmul.ac.uk}{c.mitcheltree}
\}@qmul.ac.uk, 
\href{mailto:alistair.carson@ed.ac.uk}{alistair.carson@ed.ac.uk}
}
}

\begin{document}
\ifpdf 
  \DeclareGraphicsExtensions{.png,.jpg,.pdf}
\else  
  \DeclareGraphicsExtensions{.eps}
\fi

\maketitle

\begin{abstract}
Infinite impulse response filters are an essential building block of many time-varying audio systems, such as audio effects and synthesisers. 
However, their recursive structure impedes end-to-end training of these systems using automatic differentiation. 
Although non-recursive filter approximations like frequency sampling and frame-based processing have been proposed and widely used in previous works, they cannot accurately reflect the gradient of the original system.
We alleviate this difficulty by re-expressing a time-varying all-pole filter to backpropagate the gradients through itself, so the filter implementation is not bound to the technical limitations of automatic differentiation frameworks. 
This implementation can be employed within audio systems containing filters with poles for efficient gradient evaluation. 
We demonstrate its training efficiency and expressive capabilities for modelling real-world dynamic audio systems on a phaser, time-varying subtractive synthesiser, and compressor. 
We make our code and audio samples available and provide the trained audio effect and synth models in a VST plugin\footnote{\href{https://diffapf.github.io/web/}{https://diffapf.github.io/web/}}.

\end{abstract}

\vspace{-5pt}
\section{Introduction}
\label{sec:intro}

Infinite impulse response (IIR) filters are commonly used in many time-varying audio processing units, such as subtractive synthesisers, phaser effects, and dynamic range compression.
Their recursive computation, using results from previous time steps, allows modelling a wide range of responses with low computational costs. 
Since differentiable DSP (DDSP)~\cite{Engel2020DDSP} emerged as an effective solution in attaining controllable audio systems, there have been attempts to incorporate recursive filters into automatic differentiation frameworks such as PyTorch.
However, naive implementations result in deep computational graphs due to recursion that cannot be parallelised~\cite{yu_singing_2023} and slow down training speed~\cite{nercessian_lightweight_2021, steinmetz2022style, wright2022grey}.

A common acceleration approach is to evaluate the filters in the frequency domain~\cite{nercessian_lightweight_2021} and approximate time-varying behaviour by filtering on overlapping frames in parallel~\cite{juvela2019gelp, carson2023differentiable}.
Despite their popularity, these approximations have some potential drawbacks.
Frame windowing for overlap-add smears the spectral peaks in the frequency domain, resulting in filters with artificially high resonance to counter this effect. 
Sampling the filters in the frequency domain sometimes truncates the IR length, and the circular convolution with the truncated IR caused by frequency sampling can lead to artefacts at frame boundaries.
Most importantly, systems trained in this way are not guaranteed the same results when operating sample-by-sample at audio rate to achieve low latency.

In this paper, we propose a solution to these problems by deriving and implementing an efficient backpropagation algorithm for a time-varying all-pole filter.
Our filter can be used in various audio systems by separating the system's poles and zeros and explicitly handling the poles (where the recursion is) with our proposed method fully end-to-end.
Our contributions are threefold:
\begin{enumerate}
\item We significantly increase the forward and backpropagation speed of time-varying recursive all-pole filters without introducing any approximation to the filter.
\item The systems trained with our implementation can be converted to real-time without generalisation issues besides the order of the zeros and poles.
\item We show that our filter efficiently and accurately models time-varying analog audio circuits with recursive structures.
\end{enumerate}

\vspace{-5pt}
\section{Related works}

Differentiable training of IIR filters has been explored in~\cite{bhattacharya_optimization_nodate, nercessian_neural_nodate, colonel_direct_2021, nercessian_lightweight_2021, kim_joint_2022, steinmetz2022style, carson2023differentiable}.
To sidestep the problems inherent in training over a recursion, some authors approximate IIR filters in the frequency domain using the fast Fourier transform (FFT)~\cite{nercessian_neural_nodate, colonel_direct_2021, nercessian_lightweight_2021, kim_joint_2022, steinmetz2022style}, or the short-time Fourier transform (STFT)~\cite{carson2023differentiable} for time-varying effects.
This is known as the frequency sampling (FS) method. 
It approximates the filter as time-invariant and with a finite impulse response (FIR) over the duration of a short frame, thus the accuracy of FS depends heavily on the choice of STFT parameters: the hop-size, the frame length, the FFT length, and the windowing function~\cite{SASPWEB2011}. In machine learning applications, these choices add extra hyper-parameters to models, which may require prior knowledge of the target system to set appropriately.

Time-varying all-pole filters have been used for decades in linear prediction (LP) voice synthesis~\cite{markel_linear_1976}.
Training them jointly with neural networks was first proposed in LPCNet~\cite{valin2019lpcnet}. 
The authors achieve training efficiency by using inverse-filtered speech as the target because the inverse filter has no recursion.
Other works seek to parallelise LP with frame-based processing, either filtering in the time domain~\cite{mv_sfnet_2020, yu_singing_2023} or via FS~\cite{juvela2019gelp, oh_excitglow_2020} for each frame.

Dynamic processing effects like compressors and limiters also employ time-varying recursive filters.
The filter is usually first order, and the coefficients are time-varying and dependent on the \emph{attack} or \emph{release} phases of the gain reduction signal~\cite{dafx_comp}.
In the differentiable learning context, frequency sampling can be used if the compressor's attack and release time are configured to be the same~\cite{steinmetz2022style, wright2022grey}, which simplifies the filter to be time-invariant. 
Colonel et al.~\cite{colonel2022approximating} propose dividing the gain reduction signal into attack and release passages and filtering them separately with different filters, and Guo et al.~\cite{guo22b_interspeech} downsample the signal to reduce the number of recursions.

A distinct approach that provides significant acceleration is deriving the closed-form solution of the gradients for the filter parameters and implementing it in a highly optimised way.
Bhattacharya et al.~\cite{bhattacharya_optimization_nodate} derive the instantaneous backpropagation algorithm for peak and shelving filters and train them jointly to match the response of head-related transfer functions.
Forgione et al. and Yu et al.~\cite{yu_singing_2023, forgione_dynonet_2021} decompose a time-invariant IIR filter into zeros and poles, and they show that backpropagation through an all-pole filter can be expressed using the same all-pole filter.
This method is fast because the underlying filters do not have to be implemented in an automatic differentiation framework.
Nevertheless, these solutions are made for specific filters or are only applied to time-invariant systems.

\begin{figure*}[ht]
    \centering
    \includegraphics[width=\textwidth]{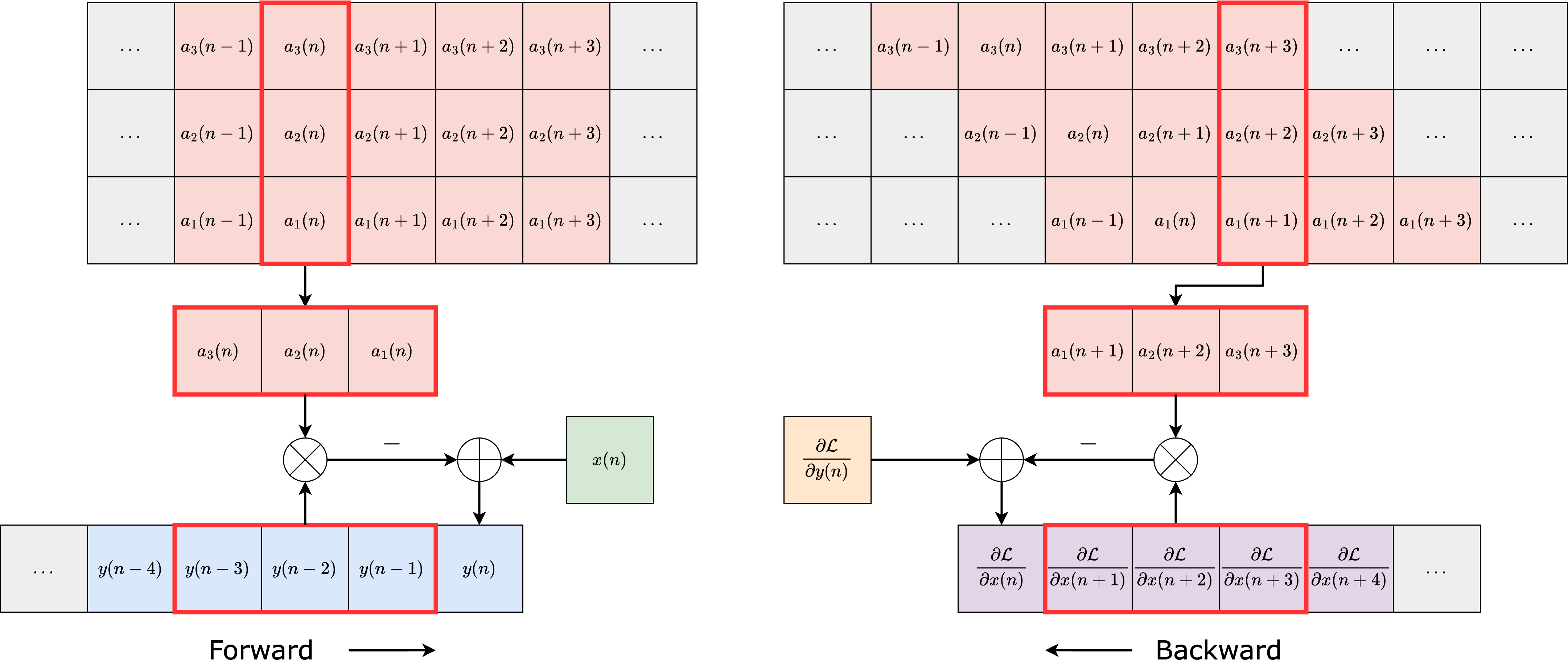}
    \caption{\itshape The forward (left) and backpropagation (right) flow chart of a third-order time-varying all-pole filter.}
    \label{fig:backprop}
    \vspace{-10pt}
\end{figure*} 

\vspace{-5pt}
\section{Proposed Methodology}
\label{sec:method}

Consider an $M^\mathrm{th}$-order time-varying all-pole filter:
\begin{equation}
\label{eq:ap}
\begin{split}
    y(n) &= f_{\mathbf{a}(n)}(x(n))\\
         &= x(n) - \sum_{i = 1}^M a_i(n) y(n-i)
\end{split}
\end{equation}
where $\mathbf{a}(n) = [a_1(n),\dots,a_M(n)]$ are filter coefficients at time $n$ and $M \in \mathbb{Z}^+$. 
In some applications, ${\bf a}(n)$ varies at a control rate $F_{c}$ much lower than the audio sampling rate $F_s$ as ${\bf a}(m)$, and can then be up-sampled to the audio rate before the filter is applied.

The following sections describe the proposed method, in which the exact gradients for each parameter of filter $f_{\mathbf{a}(n)}$ are derived and expressed in a form that can be computed efficiently.
Our method aligns most with the instantaneous backpropagation algorithms proposed in~\cite{bhattacharya_optimization_nodate, forgione_dynonet_2021, yu_singing_2023}, generalising their contributions to a time-varying all-pole filter that can be used in various recursive filters and time-varying audio systems. We refer to this method as the time domain (TD) method.

\subsection{Unwinding the Recursion}
We first rewrite the recursive Eq.~\eqref{eq:ap} so there is no $y$ variable on the right-hand side:
\begin{equation}
    y(n) = x(n) + \sum_{d=1}^\infty b_d(n) x(n-d).
    \label{eq:iir_ap}
\end{equation}
$\mathbf{b}(n) = [b_1(n), b_2(n),\dots]$ is the IIR of filter $f_{\mathbf{a}(n)}$ at time $n$.
We will use Eq.~\eqref{eq:iir_ap} and $\mathbf{b}(n)$ to help us derive the gradient of $\mathbf{a}(n)$ in the next two sections.
To get the exact value of $\mathbf{b}(n)$ in terms of $\mathbf{a}(n)$, let us think of \eqref{eq:ap} as the recursive definition of connecting any time step before $n$ to $n$.
In other words, $y(n)$ is a summation of $M$ sub-problems $y(n-i)$, each weighted by a unique coefficient $-a_i(n)$.
Let us think of $i$ as the step size going backwards from $n$ with $M$ being the maximum step size we can take.
All possible combinations of steps that span a distance $d \in \mathbb{Z}^+$ can be defined recursively as
\begin{equation}
\label{eq:G_d}
    \mathcal{G}_d = \bigcup_{i=1}^{\min(d, M)} \left\{[i; \q]: \q \in \mathcal{G}_{d-i} \right\}
\end{equation}
with boundary condition $\mathcal{G}_0 = \{[]\}$ and $[;]$ is array concatenation.

If we compare Eq.~\eqref{eq:ap} and \eqref{eq:iir_ap}, we can see that $b_d(n)$ is in fact the sum of the coefficient combinations of $a_\cdot(\cdot)$ going back from $n$ with step combinations $\mathcal{G}_d$.
Thus, the value of it is
\begin{gather}
    b_d(n) = \sum_{\q \in \mathcal{G}_d} (-1)^{|\q|} \prod_{j=1}^{|\q|} a_{q_j}\left(n - \sum_{k=1}^{j} Q(\q)_k\right) \label{eq:a2b}
\end{gather}
where $|\q|$ is the number of elements in array $\q$ and $Q(\q) = [0; \q]$.

\subsection{Gradients of $x(n)$}

Assuming we have computed $f_{\mathbf{a}(n)}$ up to step $N$, evaluated $y(n\leq N)$ with a differentiable function $\mathcal{L}(y(n))$, and have its instantaneous gradients $\frac{\partial \mathcal{L}}{\partial y(n)}$, we can backpropagate through $f_{\mathbf{a}(n)}$ as

\vspace{-10pt}
\begin{equation}
\label{eq:iir_grad_x}
\begin{split}
    \frac{\partial \mathcal{L}}{\partial x(n)} &= \sum_{d=-\infty}^{N} \frac{\partial \mathcal{L}}{\partial y(d)} \frac{\partial y(d)}{\partial x(n)}\\
    &= \frac{\partial \mathcal{L}}{\partial y(n)} \frac{\partial y(n)}{\partial x(n)} 
    + \sum_{d=1}^{N-n} \frac{\partial \mathcal{L}}{\partial y(n+d)} \frac{\partial y(n+d)}{\partial x(n)}\\
    &= \frac{\partial \mathcal{L}}{\partial y(n)} + \sum_{d=1}^{N-n} b_d(n+d) \frac{\partial \mathcal{L}}{\partial y(n+d)}.
\end{split}  
\end{equation}
We use the fact that $\frac{\partial y(n)}{\partial x(n-d)} = b_d(n)$ and $\frac{\partial y(<n)}{\partial x(n)} = 0$ from~\eqref{eq:iir_ap}.
Unfortunately, Eq.~\eqref{eq:a2b} and therefore~\eqref{eq:iir_grad_x} are expensive to compute.
We aim to express backpropagation using $f_{\mathbf{a}(n)}$, which is much more efficient to compute due to its recursion.
If we can re-parameterise Eq.~\eqref{eq:iir_grad_x} to look like \eqref{eq:ap}, then $f_{\mathbf{a}(n)}$ can be reused to compute $\frac{\partial \mathcal{L}}{\partial x(n)}$.
We do this by writing $b_d(n+d)$ in terms of $a_\cdot(\cdot)$ which requires us to evaluate~\eqref{eq:a2b} at $n+d$ to get
\begin{equation}
\label{eq:b2a}
\begin{split}
    b_d(n+d) = \sum_{\q \in \mathcal{G}_d} (-1)^{|\q|} \prod_{j=1}^{|\q|} a_{q_j}\left(n+d- \sum_{k=1}^{j} Q(\q)_k\right) \\
             = \sum_{\q \in \mathcal{G}_d} (-1)^{|\q|} \prod_{j=1}^{|\q|} a_{q_j}\left(n+ \sum_{k=j}^{|\q|} q_k\right) \\
             = \sum_{\q \in \mathcal{G}_d} (-1)^{|\q|} \hat{a}_{q_{|\q|}}(n) \prod_{j=1}^{|\q|-1} \hat{a}_{q_j}\left(n+ \sum_{k=j+1}^{|\q|} q_k\right) \\
             = \sum_{\Tilde{\q} \in \mathcal{G}_d} (-1)^{|\Tilde{\q}|} \prod_{j=1}^{|\Tilde{\q}|} \hat{a}_{\Tilde{q}_j}\left(n+ \sum_{k=1}^{j} Q(\Tilde{\q})_k\right) \\
\end{split}
\end{equation}
where $\Tilde{\q} = [q_{|\q|},\ldots,q_1] \in \mathcal{G}_d$ and $\hat{a}_i(n) = a_i(n+i)$.
Eq.~\eqref{eq:b2a} is now the same as~\eqref{eq:a2b} but uses $\hat{\mathbf{a}}(n) = [\hat{a}_1(n),\ldots,\hat{a}_M(n)]$ as coefficients and the plus sign inside the product changes to a minus sign, which means the filter should be applied in the reverse direction ($n = N \rightarrow n = -\infty$) and is supported by the non-causal indexing in~\eqref{eq:iir_grad_x} ($n+d$ instead of $n-d$). 
We can now express~\eqref{eq:iir_grad_x} in terms of $f_{\mathbf{a}(n)}$ using~\eqref{eq:b2a} and the equivalence between $\mathbf{a}(n)$ and $\mathbf{b}(n)$ from~\eqref{eq:ap} and~\eqref{eq:iir_ap} to get
\begin{equation}
\label{eq:grad_x}
\begin{split}
    \frac{\partial \mathcal{L}}{\partial x(n)}
    &= \frac{\partial \mathcal{L}}{\partial y(n)} - \sum_{i = 1}^M \hat{a}_i(n) \frac{\partial \mathcal{L}}{\partial x(n+i)} \\
    &= \text{FLIP} 
        \circ f_{\text{FLIP} \circ \hat{\bf a}(n)}
        \circ \text{FLIP}
        \circ \frac{\partial \mathcal{L}}{\partial y(n)}
\end{split}
\end{equation}
where $\text{FLIP}(x(n)) = x(-n)$ and $f_1 \circ f_2(x) = f_1(f_2(x))$. 
$\text{FLIP}$ and $\hat{\bf a}(n)$ are trivial to compute using memory indexing.
The backpropagation algorithm and how to arrange $\mathbf{a}(n)$ into $\hat{\mathbf{a}}(n)$ is shown in Figure~\ref{fig:backprop}.

\subsection{Gradients of $\mathbf{a}(n)$}

Let $u_i(n) = -a_i(n)y(n-i)$ so $y(n) = x(n) + u_1(n) + \cdots + u_M(n)$.
Because of the chain rule and $\frac{\partial y(n)}{\partial x(n)} = \frac{\partial y(n)}{\partial u_i(n)} = 1$, $x(n)$ and $u_i(n)$ should have the same derivatives ($\frac{\partial \mathcal{L}}{\partial x(n)} = \frac{\partial \mathcal{L}}{\partial u_i(n)}$).
Since we can compute $\frac{\partial \mathcal{L}}{\partial x(n)}$ from~\eqref{eq:grad_x}, the gradients of the coefficients are simply
\begin{equation}
\label{eq:grad_a}
    \frac{\partial \mathcal{L}}{\partial a_i(n)} 
    = \frac{\partial \mathcal{L}}{\partial u_i(n)} \frac{\partial u_i(n)}{\partial a_i(n)}
    = -\frac{\partial \mathcal{L}}{\partial x(n)} y(n-i).
\end{equation}

\noindent In summary, we can calculate all gradients with one pass of $f_{\mathbf{a}(n)}$ and the multiplications in~\eqref{eq:grad_a}, which are fast to compute.
We implement an efficient $f_{\mathbf{a}(n)}$ for both forward and backward computation using Numba and register it as a custom operator in PyTorch.
The implementation is available on GitHub~\footnote{\href{https://github.com/DiffAPF/torchlpc}{https://github.com/DiffAPF/torchlpc}}.

\vspace{-5pt}
\section{Applications}

We demonstrate our all-pole filter implementation on three commonly used dynamic audio systems: a phaser, a subtractive synthesiser, and a compressor. 
All three systems have time-varying recursive structures that are not easy to train in a differentiable way and would typically be modelled using FS approaches.

Although the filtering order of poles and zeros matters in time-varying systems, in this work, we rearranged the poles of each system into one all-pole filter for maximum training efficiency. 
Due to the relatively slowly varying filter coefficients, we found this approach to be sufficient.
We direct interested readers to our parallel work in speech synthesis~\cite{ycy2024golf}, an exact all-pole system.

\subsection{Phaser}\label{sec:apps:subsec:phaser}
\begin{figure}[h!]
    \vspace{-10pt}
    \centering
    \includegraphics[width=.49\textwidth]{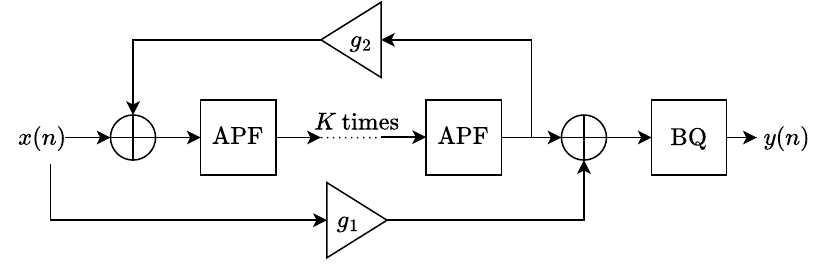}
    \caption{\it Discrete-time phaser model considered in this work, where $K=4$. APF represents a time-varying all-pass filter with difference equation \eqref{eq:apf} and BQ is a biquad filter.}
    \label{fig:phaser}
\end{figure}

We test our filter implementation on a virtual analog phaser model, based on~\cite{PASPWEB2010, carson2023differentiable}. At the core of the model is a differentiable LFO that operates at the control rate $F_c$. The oscillator is implemented as a damped oscillator with learnable frequency $f_0$, decay rate $\sigma$, and phase $\phi$:
\begin{equation}\label{eq:osc}
    s(m) = e^{-\sigma^2m/F_c}\cos(2\pi f_0m/F_c + \phi)
\end{equation}
where $m$ is the control-rate sample index. The inclusion of parameter $\sigma$ alleviates some non-convexity issues when learning the frequency $f_0$, as shown in~\cite{Hayes2022:sine}. Note that the oscillator is unconditionally stable for all $\sigma$. The oscillator is passed through a multi-layer perceptron (MLP) network to obtain the control signal $p(m)$. The MLP, with parameters $\Theta$, contains 3x8 hidden layers, with $\tanh$ activation functions on all layers including the final. The control signal is then up-sampled with linear interpolation to obtain $p(n)$ and modulates the coefficients of four cascaded first-order all-pass filters (APF), each with the difference equation:
\begin{equation}\label{eq:apf}
    y_k(n) = p(n) \cdot \left[ x_k(n) + y_k(n-1) \right] - x_k(n-1)
\end{equation}
where $x_k$ and $y_k$ are the input and state of the $k^\mathrm{th}$ APF, respectively, $0 \leq k < 4$. Note that the MLP $\tanh$ output activation ensures the all-pass poles remain within the unit circle. The APFs are arranged in series, with a through path of gain $g_1$ and feedback loop $g_2$ as shown in Figure \ref{fig:phaser}.  It is common to include a unit delay in the feedback path for ease of implementation~\cite{PASPWEB2010, Kiiski2016}, however here we use instantaneous feedback for a more realistic virtual analog model. In Figure \ref{fig:phaser}, BQ represents a biquad filter with coefficients ${\bf b}^{({\rm bq})} = [b_0^{({\rm bq})}, b_1^{({\rm bq})}, b_2^{({\rm bq})}]$ and ${\bf a}^{({\rm bq})} = [a_1^{({\rm bq})}, a_2^{({\rm bq})}]$. The entire model is a sixth-order time-varying IIR filter. Here we approximate the system as having the difference equation:
\begin{align}\label{eq:phaser}
    y(n) = f_{{\bf a}(n)} \circ f_{{\bf b}(n)} \circ x(n)
\end{align}
where $f_{{\bf b}(n)}\left(\cdot\right)$ is a time-varying FIR filter:
\begin{equation}\label{eq:phaser_b}
    f_{{\bf b}(n)}\left(x(n)\right) = \sum_{i=0}^{M}b_i(n)x(n-i)
\end{equation}
and $f_{{\bf a}(n)}\left(\cdot\right)$ is a time-varying all-pole filter (see Eq. \eqref{eq:ap}). 
Here $M=6$ for $\mathbf{b}(n)$ and $\mathbf{a}(n)$, which are functions of $p(n)$, $g_1$, $g_2$, ${\bf b}^{({\rm bq})}$, and ${\bf a}^{({\rm bq})}$. 
In previous work~\cite{carson2023differentiable}, the control parameters of a similar time-varying filter were learned through gradient descent using the FS method. This frequency sampling approach had some limitations, however. Firstly, the optimal frame size for the best training accuracy depended on the rate of the target LFO, which we ideally should not assume as prior knowledge. Secondly, it was not fully investigated whether the trained model could then be implemented in the time domain at inference to avoid latency.

Here, we instead implement Eq. \eqref{eq:phaser} directly in the time domain during training using the method proposed in Section \ref{sec:method}.

\vspace{-15pt}
\subsection{Time-varying Subtractive synthesiser}\label{sec:apps:subsec:synth}

We test our filter implementation on a subtractive synthesiser roughly modelled after the \emph{Roland TB-303 Bass Line} synth\footnote{\href{https://www.roland.com/uk/promos/303day/}{https://www.roland.com/uk/promos/303day/}} which defined the acid house electronic music movement of the late 1980s.
The TB-303 is an ideal synth for our use case because its defining feature is a resonant low-pass filter where the cutoff frequency is modulated quickly using an envelope to create its signature squelchy,\\``liquid'' sound.
Although the original TB-303's circuit contains a 4-pole diode ladder filter, for simplicity and demonstration purposes, we implement our synthesiser using a biquad filter.
Our synth is differentiable and consists of three main components: a monophonic oscillator, a time-varying biquad filter, and a waveshaper for adding distortion to the output.

The oscillator is the same as in the one in \texttt{TorchSynth}~\cite{turian2021torchsynth} and uses hyperbolic tangent waveshaping to generate sawtooth or square waves, and can sweep continuously between them. It is defined by the following equations:

\vspace{-10pt}
\begin{gather}
    \psi(n) = 2\pi n f_0 / F_s + \phi \pmod{2\pi} \\
    o(n) = \rho_{\rm osc} s_{\rm saw}(\psi(n)) + (1 - \rho_{\rm osc}) s_{\rm sq}(\psi(n)) \\
    e(n) =
    \begin{rcases}
        \begin{dcases}
            \left( \frac{N_{\rm on} - n}{N_{\rm on}} \right)^{\rho_{\rm env}} &  0 \leq n \leq N_{\rm on} \\
            \omit \hfil 0 \hfil &  {\rm otherwise} \\
        \end{dcases}
    \end{rcases} \\
    s(n) = g_{\rm osc} e(n) o(n)\label{eq:synth_osc}
\end{gather}

\noindent where $F_s$ and $f_0$ are the sampling rate and fundamental frequency in hertz, and $\phi$ is the phase in radians. $\rho_{\rm osc}$ is a continuous control parameter to sweep between the wave shapes where 0 makes a square wave ($s_{\rm sq}(\cdot)$), and 1 makes a saw wave ($s_{\rm saw}(\cdot)$). 
The output audio of the oscillator is multiplied by gain $g_{\rm osc}$ and is then shaped using a decaying envelope $e(n)$ of length $N_{\rm on}$ note on samples with control parameter $\rho_{\rm env}$.

A time-varying biquad filter $h_{\rm bq}(\cdot)$ (same as Equations \ref{eq:phaser} and \ref{eq:phaser_b} for the phaser, but $M = 2$) is then applied to the oscillator output audio. 
This filter takes as input 5 time-varying filter coefficients \{$a_1(n)$, $a_2(n)$, $b_0(n)$, $b_1(n)$, $b_2(n)$\} at sample rate which can be passed in directly or generated from filter cutoff and resonance modulation signals $m_{\rm fc}(n)$ and $m_{\rm q}(n)$. 
These modulation signals are then used to calculate the coefficients for a biquad lowpass filter using the corresponding equations in the Audio EQ Cookbook~\footnote{\href{https://www.w3.org/TR/audio-eq-cookbook/}{https://www.w3.org/TR/audio-eq-cookbook/}}.

Finally, the output of the filter is fed through a hyperbolic tangent waveshaper which adds distortion:

\begin{equation}
\label{eq:synth_dist}
    f_{\rm{dist}}(x(n)) = \tanh(g_{\rm dist} x(n)).
\end{equation}

\noindent The amount of distortion is controlled by parameter $g_{\rm dist}$ which modifies the gain of the input of the waveshaper $x(n)$.

The entire synth is therefore controllable using 8 global parameters $\{f_0$, $F_s$, $\phi$, $N_{\rm on}$, $\rho_{\rm osc}$, $\rho_{\rm env}$, $g_{\rm osc}$, $g_{\rm dist}\}$ and 2 or 5 time-varying parameters $\{m_{\rm fc}(n)$, $m_{\rm q}(n)\}$ or $\{a_1(n)$, $a_2(n)$, $b_0(n)$, $b_1(n)$, $b_2(n)\}$.
It is defined by composing Equation~\ref{eq:synth_osc}, $h_{\rm bq}(\cdot)$, and Equation~\ref{eq:synth_dist} together as follows:

\begin{equation}
    y_{\rm synth}(n) = f_{\rm{dist}} \circ h_{\rm bq} \circ s(n).
\end{equation}

\subsection{Feed-forward Compressor}
\label{ssec:ff_comp}

The compressor we consider here is a simple feed-forward compressor from~\cite{dafx_comp}, which is defined as:

\begin{gather}
x_{\rm rms}(n) = \alpha_{\rm rms}x^2(n) + (1 - \alpha_{\rm rms})x_{\rm rms}(n-1) \label{eq:rms}\\
g(n) = \min \left(1, \left(\frac{\sqrt{x_{\rm rms}(n)}}{10^{\frac{CT}{20}}} \right)^{\frac{1-R}{R}} \right)\\ 
\hat{g}(n) =  
\begin{rcases}
    \begin{dcases}
        \alpha_{\rm at}g(n) + (1 - \alpha_{\rm at})\hat{g}(n-1) &  g(n) < \hat{g}(n-1) \\
        \alpha_{\rm rt}g(n) + (1 - \alpha_{\rm rt})\hat{g}(n-1) &  {\rm otherwise} \\
    \end{dcases}
\end{rcases}\label{eq:dyn_filter}\\
y(n) = x(n)\hat{g}(n)\gamma.
\end{gather}
$R$, $CT$, and $\gamma$ are the ratio, threshold, and make-up gain, respectively.
$\alpha_{\rm rms/at/rt}$ are the average smoothing coefficients.
$\alpha_{\rm at/rt}$ are chosen based on whether the compressor is operated in the \emph{attack} phase or \emph{release} phase.
This compressor's training efficiency bottleneck is described in~\eqref{eq:dyn_filter}, as the coefficient of a recursive filter is computed \emph{on the fly}, so we cannot use $f_{\mathbf{a}(n)}$ directly.
Moreover, it operates at the audio rate, so it is unsuitable for frame-based approximation.

\begin{table*}
\centering
\caption{\itshape Parameter limits in our differentiable DSP models.}
\vspace{-5pt}
\resizebox{\textwidth}{!}{
\begin{tabular}{cccccccccccccccccc}
    \toprule
    & \multicolumn{5}{c}{Phaser} & \multicolumn{6}{c}{Time-varying subtractive synth} & \multicolumn{6}{c}{Feed-forward compressor} \\
    \cmidrule(lr){2-6} \cmidrule(lr){7-12} \cmidrule(lr){13-18}
    & $f_0$ & $\sigma$ & $\phi$ & $g_1$ & $g_2$ & $\rho_{\rm osc}$ & $\rho_{\rm env}$ & $g_{\rm osc}$ & $g_{\rm dist}$ & $m_{\rm fc}(n)$ & $m_{\rm q}(n)$ & $R$ & $CT$ & $\alpha_{\rm at}$ & $\alpha_{\rm rt}$ & $\alpha_{\rm rms}$ & $\gamma$ \\ 
    \cmidrule(lr){2-6} \cmidrule(lr){7-12} \cmidrule(lr){13-18}
    min. & $-F_c/2$ & $-\infty $  & $-\pi$ & $- \infty $ & 0 & 0.0 & 0.1 & 0.01 & 0.01 & \SI{0.1}{kHz} & 0.7071 & 1.0    & $-\infty$ & 0.0 & 0.0 & 0.0 & 0.0 \\
    max. & $F_c/2$ & $\infty$ & $\pi$ & $ \infty $ & 1 & 1.0 & 6.0 & 1.00 & 4.00 & \SI{8.0}{kHz} & 8.0 & $\infty$ & $\infty$ & 1.0 & 1.0 & 1.0 & $\infty$ \\
    \bottomrule
\end{tabular}
}
\label{tab:all_params}
\vspace{-10pt}
\end{table*}

\subsubsection{Custom backward function}
We backpropagate gradients through~\eqref{eq:dyn_filter} using the proposed method in Sec.~\ref{sec:method}.
To handle the if-else statement, we write the gain reduction filter~\eqref{eq:dyn_filter} in Numba and record each if-else decision inside the recursion into a binary mask $\zeta(n)$:
\begin{equation}
\label{eq:dyn_mask}
\zeta(n) =
\begin{rcases}
    \begin{dcases}
        1 &  g(n) < \hat{g}(n-1) \\
        0 &  {\rm otherwise} \\
    \end{dcases}
\end{rcases}.
\end{equation}
Using this, Eq.~\eqref{eq:dyn_filter} equals the following time-varying IIR:
\begin{equation}
\label{eq:dyn_filter2}
\begin{split}
\beta(n) &= \alpha_{\rm at}^{\zeta(n)}\alpha_{\rm rt}^{1-\zeta(n)}\\
\hat{g}(n) &= \beta(n) g(n) + (1 - \beta(n))\hat{g}(n-1).
\end{split}
\end{equation}

\pagebreak

We can now use $\beta(n)$ and $f_{\mathbf{a}(n)}$ to backpropagate the gradients through the compressor.
The pseudo-code of the differentiable gain reduction filter is summarised in Algorithm~\ref{alg:dyn}, and the implementation can be found on GitHub~\footnote{\href{https://github.com/DiffAPF/torchcomp}{https://github.com/DiffAPF/torchcomp}}.

\vspace{-5pt}
\begin{algorithm}
\caption{Differentiable gain reduction filter~\eqref{eq:dyn_filter}.}\label{alg:dyn}
  \DontPrintSemicolon
  \SetKwFunction{FForward}{forward}
  \SetKwFunction{FBackward}{backward}
  \SetKwProg{Fn}{def\,}{:}{}
  \Fn{\FForward{$g$, $\alpha_{\rm at}$, $\alpha_{\rm rt}$}}{
        $\hat{g}(-1) \leftarrow 1$\;
        $\hat{g}(n) \leftarrow$ Eq.~\eqref{eq:dyn_filter}\;
        $\zeta(n) \leftarrow$ Eq.~\eqref{eq:dyn_mask} \;
        \KwRet $\hat{g}$, $\zeta$\;
  }
  \;
  \Fn{\FBackward{$\frac{\partial \mathcal{L}}{\partial \hat{g}}$,
                $\hat{g}$, $g$, $m$, $\alpha_{\rm at}$, $\alpha_{\rm rt}$}}{
        $\beta(n) \leftarrow$ Eq.~\eqref{eq:dyn_filter2}\;
        $\frac{\partial \mathcal{L}}{\partial \beta(n)g(n)} \leftarrow$ filtering $\frac{\partial \mathcal{L}}{\partial \hat{g}(n)}$ with Eq.~\eqref{eq:grad_x} and $a_1(n) \equiv \beta(n) - 1$\;
        $\frac{\partial \mathcal{L}}{\partial g(n)} \leftarrow \frac{\partial \mathcal{L}}{\partial \beta(n)g(n)} \beta(n)$\;
        $\frac{\partial \mathcal{L}}{\partial \beta(n)} \leftarrow 
            \frac{\partial \mathcal{L}}{\partial \beta(n)g(n)} (g(n) - \hat{g}(n-1))$\;
        $\frac{\partial \mathcal{L}}{\partial \alpha_{\rm at}} \leftarrow \sum_n \frac{\partial \mathcal{L}}{\partial \beta(n)} \zeta(n)$\;
        $\frac{\partial \mathcal{L}}{\partial \alpha_{\rm rt}} \leftarrow \sum_n \frac{\partial \mathcal{L}}{\partial \beta(n)} (1 - \zeta(n))$\;
        \KwRet $\frac{\partial \mathcal{L}}{\partial g}$, 
               $\frac{\partial \mathcal{L}}{\partial \alpha_{\rm at}}$,
               $\frac{\partial \mathcal{L}}{\partial \alpha_{\rm rt}}$\;
  }
\end{algorithm}
\vspace{-8pt}

\vspace{-5pt}
\section{Experiments}

All three systems are trained to model some target analog audio in an end-to-end fashion using gradient descent.
We use the Hanning window for all the frame-based approaches.
The ranges of all the interpretable parameters are summarised in Table~\ref{tab:all_params}.
We provide audio samples for all experiments on the accompanying website.

\subsection{Modelling the EHX Small Stone analog phaser}\label{sec:exp:subsec:phaser}

Here we explore using the DSP phaser model outlined in Sec. \ref{sec:apps:subsec:phaser} to model an analog phaser pedal: the Electro-Harmonix Small-Stone. This is the same system modelled in~\cite{carson2023differentiable} using the FS method. The circuit consists of four cascaded analog all-pass filters, a through-path for the input signal, and a feedback path~\cite{jdsleep} -- so topologically the circuit is similar to the discrete-time phaser model considered in this paper. The pedal consists of one knob which controls the LFO rate, and a switch that engages the feedback loop. Six different parameter configurations are considered:
\begin{itemize}
    \item SS-A:   feedback off, rate knob 3 o'clock ($f_0 \approx$  \SI{2.3}{\hertz}) 
    \item SS-B:   feedback off, rate knob 12 o'clock ($f_0 \approx$  \SI{0.6}{\hertz})   
    \item SS-C:   feedback off, rate knob 9 o'clock ($f_0 \approx$  \SI{0.09}{\hertz}) 
    \item SS-D:   feedback on,  rate knob 3 o'clock ($f_0 \approx$  \SI{1.4}{\hertz})
    \item SS-E:   feedback on,  rate knob 12 o'clock ($f_0 \approx$  \SI{0.4}{\hertz}) 
    \item SS-F:   feedback on,  rate knob 9 o'clock ($f_0 \approx$  \SI{0.06}{\hertz}) 
\end{itemize}

\begin{table}[h]
  \centering
  \caption{\it Phaser evaluation results. ``Method'' refers to the training method; whereas the ESR was computed over the test dataset using both FS and TD at inference. }
  \vspace{-5pt}
  \label{tab:phaser_eval}
  \resizebox{0.65\columnwidth}{!}{
    \begin{tabular}{llccc}
    \toprule
    \multirow{2}{*}{Dataset} & \multirow{2}{*}{$L/F_s$} & \multirow{2}{*}{Method} & \multicolumn{2}{c}{ESR (\%)} \\
    \cmidrule{4-5}
    & & & FS & TD \\
    \midrule
    \multirow{2}{*}{SS-A} & \multirow{2}{*}{\SI{10}{\ms}} & FS & 1.46  & 1.53  \\
    & & TD &   \bf 1.34  & \bf 1.36  \\
    \midrule
    \multirow{2}{*}{SS-B}  & \multirow{2}{*}{\SI{40}{\ms}} & FS & 1.37  & 1.49  \\
    & & TD & \bf 1.35  & \bf 1.34  \\
    \midrule
    \multirow{2}{*}{SS-C}  & \multirow{2}{*}{\SI{160}{\ms}} & FS & \bf 1.62  & \bf 1.80  \\
    & & TD & 2.56  &  2.23  \\
    \midrule
    \multirow{2}{*}{SS-D}  & \multirow{2}{*}{\SI{10}{\ms}} & FS & 22.47  & 23.47  \\
    & & TD &  \bf 21.64  & \bf 23.33  \\
    \midrule
    \multirow{2}{*}{SS-E}  & \multirow{2}{*}{\SI{40}{\ms}}  & FS & 15.43  & 16.69  \\
    & & TD & \bf 13.63  & \bf 13.87  \\
    \midrule
    \multirow{2}{*}{SS-F} & \multirow{2}{*}{\SI{160}{\ms}}  & FS & 8.79   & 9.83  \\
    & & TD & \bf 7.83  & \bf 8.79  \\
    \bottomrule
    \end{tabular}%
  }
  \label{tab:addlabel}%
  \vspace{-15pt}
\end{table}%

The training data consists of a \SI{30}{\second} chirp-train both dry (input) and processed through the pedal (target). At each training iteration, the input signal is processed through the model in a single batch, and the loss function is computed as the error-to-signal ratio (ESR) between the model output and target. The learnable model parameters are $\{g_1, g_2, f_0, \sigma, \phi, \Theta,  {\bf b}^{({\rm bq})}, {\bf a}^{({\rm bq})}\}$, as defined in Sec. \ref{sec:apps:subsec:phaser}, giving a total of 182 model parameters. An Adam optimiser with a learning rate $5\times 10^{-4}$ is employed to carry out parameter updates every iteration for a maximum of 10k iterations. The test data includes the training data plus the next 10 seconds of the same audio signal, which contains guitar playing. This ensures the learned LFO phase is always aligned to the same point in time.

As reported in~\cite{carson2023differentiable}, the accuracy and convergence of model training depends on the choice of hop-size and window-length. 
Furthermore, even for a fixed choice of hyper-parameters, the training convergence depends on the initial parameter values, which are pseudo-randomly initialised~\cite{carson2023differentiable}.
We observed that for some random seeds, the LFO would not converge to the correct frequency $f_0$ and/or decay rate $\sigma$. In a successful run, the learned $f_0$ is approximately equal to that of the target, and $\sigma$ converges to zero. 

As a baseline, we train the model using the FS method with a single hop-size $L$ for each parameter configuration. The window-length $N_{\rm WIN}$ is set to four times the hop-size, and the FFT length to $2^{\lceil \log_2(N_{\rm WIN})\rceil + 1}$. The training process is repeated up to five times with different seeds until a model converges. The proposed time domain (TD) implementation is then trained with the same hyper-parameters (where relevant) and initial seed as the FS method. Here, the hop-size determines the control rate. 

To evaluate the two methods, we train the phaser model with the respective method and then test using both FS and TD at inference time. The test ESR can be seen in Table \ref{tab:phaser_eval}. For all datasets, it can be seen that both methods result in a very similar test loss, with the TD method doing slightly better in five out of the six datasets. Of course, only one hop-size has been considered here, so a more detailed analysis across a range of hop-sizes would be an interesting area of further work. However, these specific hop-sizes were chosen based on the recommendations in our previous work~\cite{carson2023differentiable}, in which they were heuristically found to give the best results (for the respective datasets) using the FS training method. 

In early experiments we observed that the TD implementation could become unstable during training, causing the output signal to explode. This occurred even for a simplified problem of $g_2 = 0$ and with the exclusion of the BQ filter. It was verified that the APF poles were within the unit circle for all $n$, albeit close in some cases ($p(n)>0.98$). We, therefore, suspect this instability was due to numerical inaccuracies associated with the transient response of the filters (e.g. as described in~\cite{SASPWEB2011}) because changing from single-precision to double-precision resolved this problem. Therefore, we recommend operating at double precision when using the proposed filter implementation if instability arises.

\subsection{Modelling the Roland TB-303 Acid Synth}

We model analog TB-303 audio with our time-varying subtractive synth.
The dataset is made from Sample Science's royalty free \emph{Abstract 303} sample pack\footnote{\href{https://www.samplescience.info/2022/05/abstract-303.html}{https://www.samplescience.info/2022/05/abstract-303.html}} consisting of 100 synth loops at 120~BPM recorded dry from a hardware TB-303 clone.
All loops are concatenated together, resampled to \SI{48}{\kHz}, and then pitch and note on durations are extracted using Ableton Live 11's melody-to-midi conversion algorithm which we found to be highly accurate for these monophonic melody loops.
Since the TB-303 is a 16 note sequencer, the resulting annotated notes are truncated or zero-padded to 6000 samples (one 16th note at 120~BPM and \SI{48}{\kHz}) with any notes shorter than 4000 samples in duration thrown out. 
This is then split into 60\%, 20\%, and 20\% train, validation, and test sets, respectively, resulting in a total of 42.5 seconds of audio.

\begin{figure}[h!]
    \centering
    \vspace{-10pt}
    \includegraphics[width=0.96\columnwidth]{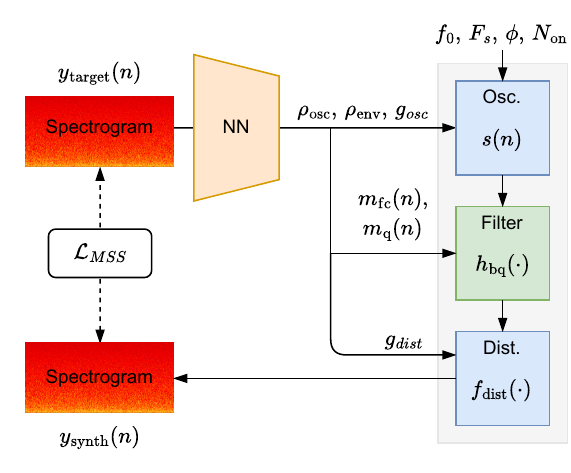}
    \vspace{-15pt}
    \caption{\it Diagram of the differentiable synth modelling process. Our time domain filter component is shown in green.}
    \label{fig:synth}
    \vspace{-11pt}
\end{figure}

We use the same modulation extraction approach as~\cite{mitcheltree2023modulation} and DDSP~\cite{Engel2020DDSP} to model the synth. 
First, frame-by-frame features are extracted from the target audio and are processed by a neural network which predicts the temporal and global control parameters for our differentiable synth as defined in Section~\ref{sec:apps:subsec:synth}.
Temporal parameters are linearly interpolated from frame-rate to sample-rate when required.
The synth then generates some reconstructed audio from the control parameters which can be compared against the target audio using a differentiable loss function, thus enabling the system to be trained end-to-end using gradient descent.
A diagram of the entire setup is shown in Figure~\ref{fig:synth}.

Since the filter modulations of the TB-303 are very fast (around \SI{125}{\ms} for 16th notes at 120~BPM), we use a Mel spectrogram with 1024 FFT size, 128 Mel bins, and a short hop length of 32 samples which results in 188 frames for 6000 samples of audio. 
The neural network is the same architecture as LFO-net~\cite{mitcheltree2023modulation} except with 4 or 5 additional 2-layer MLP networks applied to the latent embedding averaged across the temporal axis to predict the global parameters of the synth.
It contains 730~K parameters. 
We conduct experiments with five different synth filter configurations:

\begin{enumerate}
    \item Time domain biquad coefficients (Coeff TD) 
    \item Frequency sampling biquad coefficients (Coeff FS)
    \item Time domain low-pass biquad (LP TD)
    \item Frequency sampling low-pass biquad (LP FS)
    \item Time domain recurrent neural network (LSTM)
\end{enumerate}

\noindent As discussed in Section~\ref{sec:apps:subsec:synth}, for configurations 1 and 2 the neural network outputs a 5-dimensional temporal control parameter of changing biquad coefficients. 
Before filtering, these coefficients are post-processed using the biquad triangle parameterisation of~\cite{nercessian_lightweight_2021} to ensure stability. 
For configurations 3 and 4, a 2-dimensional modulation signal is returned, representing a changing filter cutoff and its Q factor (which is constant over time).
These are then converted to five biquad coefficients.
The raw coefficient filter configuration gives the synth as many degrees of freedom as possible whereas the lowpass filter configuration is based on the TB-303's analog design consisting of an envelope modulated lowpass filter with a global resonance control knob.
Finally, the recurrent neural network filter (configuration 5) is based on the architecture in~\cite{wright2022grey, mitcheltree2023modulation} and enables us to compare against learning a time-varying IIR filter directly from scratch.
It is conditioned with the same 2-dimensional modulation signal as configurations 3 and 4.

We train all models for 200 epochs on batches of 34 notes using the AdamW optimiser and single precision -- the numerical issues noted in Sec. \ref{sec:exp:subsec:phaser} were not observed here. 
For the FS synth filter configurations we train separate models for $N_{\rm WIN} \in [128, 256, 512, 1024, 2048, 4096]$ while keeping the hop-size $L$ fixed at 32 samples to match the frame-rate temporal outputs of the modulation extraction neural network. 
As in the phaser experiments, the FFT length is set to $2^{\lceil \log_2(N_{\rm WIN})\rceil + 1}$.
The LSTM configuration is trained with 64 hidden units.
Note pitch and durations are provided to the synth for reconstruction and the phase for the oscillator is random to improve robustness and reflect how synths behave in reality.
As a result, the target and reconstructed audio may be misaligned which is why we use multi-resolution STFT loss (MSS)~\cite{yamamoto2020parallel} for training which is phase agnostic.

We evaluate the different filter configurations by comparing their MSS loss values on the 20\% test split.
The FS configurations that operate at frame-rate are also evaluated at sample-rate by linearly interpolating their filter coefficients during inference.
We also calculate the Fréchet Audio Distance (FAD)~\cite{kilgour19_interspeech} for each model which has been shown to correlate with human perception.
Since the individual audio files are very short, we first concatenate them into one audio file before calculating the FAD.
To avoid harsh discontinuities in the concatenated file, we apply a 32-sample fade (one hop-size $L$) to both ends of individual clips.
The evaluation results are summarised in Table~\ref{tab:synth_eval}.
Finally, in Table~\ref{tab:synth_bench} we show the speed benchmarks of the different synth configurations.

\begin{table}[h]
\centering
\caption{\itshape Synth evaluation results (FS = frequency sampling, TD = time domain) with 95\% confidence intervals for FAD scores.}
\vspace{-5pt}
\resizebox{\columnwidth}{!}{
\begin{tabular}{ccccccc}
    \toprule
    &&& \multicolumn{2}{c}{MSS} & \multicolumn{2}{c}{FAD VGGish} \\
    \cmidrule(lr){4-5} \cmidrule(lr){6-7}
    Filter & Method & $N_{\rm WIN}$ & FS & TD & FS & TD \\ 
    \midrule
    \multirow{7}{*}{Coeff.} & \multirow{6}{*}{FS} &   4096    & 1.66 & 1.78 & 2.62 $\pm$ 0.09 & 2.70 $\pm$ 0.13 \\
     &  & 2048  & 1.64 & 1.65 & \bf 2.18 $\pm$ 0.07 & 2.35 $\pm$ 0.11 \\
     &  & 1024  & 1.53 & 1.58 & 2.57 $\pm$ 0.08 & 2.27 $\pm$ 0.12 \\
     &  & 512   & 1.57 & 1.57 & 2.87 $\pm$ 0.10 & 2.46 $\pm$ 0.10 \\
     &  & 256   & \bf 1.49 & 1.48 & 2.25 $\pm$ 0.08 & \bf 1.98 $\pm$ 0.06 \\
     &  & 128   & 1.53 & 1.55 & 3.37 $\pm$ 0.14 & 2.73 $\pm$ 0.12 \\
     \cmidrule{2-7}
      &TD&  - & -    & \bf 1.38 & -    & 2.49 $\pm$ 0.21 \\
    \midrule
    \multirow{7}{*}{LP} & \multirow{6}{*}{FS} & 4096    & 1.96 & 1.98 & 2.59 $\pm$ 0.06 & \bf 2.09 $\pm$ 0.07 \\
     &  & 2048    & 1.95 & 2.04 & 2.62 $\pm$ 0.07 & 4.52 $\pm$ 0.17 \\
     &  & 1024    & 1.89 & 2.15 & 2.59 $\pm$ 0.08 & 4.18 $\pm$ 0.14 \\
     &  & 512     & 1.83 & 2.92 & \bf 2.13 $\pm$ 0.06 & 3.38 $\pm$ 0.08 \\
     &  & 256     & \bf 1.82 & 2.89 & 2.17 $\pm$ 0.06 & 3.36 $\pm$ 0.12 \\
     &  & 128     & 1.84 & 2.70 & 2.34 $\pm$ 0.09 & 3.93 $\pm$ 0.12 \\
     \cmidrule{2-7}
     & TD  &-&-& \bf 1.56 & -    & 2.51 $\pm$ 0.10 \\
    \midrule
    LSTM 64 & TD &-     & -    & 1.76 & -    & 3.24 $\pm$ 0.07 \\
    \bottomrule
\end{tabular}
}
\label{tab:synth_eval}
\vspace{-5pt}
\end{table}

\begin{table}
    \centering
    \caption{\itshape Synth CPU runtime benchmarks on an M1 Pro MacBook for one optimisation step (forward + backward, one thread, single precision, batch size of 34).}
    \vspace{-5pt}
    \resizebox{1\columnwidth}{!}{
    \begin{tabular}{cccccccc}
        \toprule
        \multirow{2}{*}{Synth} & \multirow{2}{*}{TD} & \multicolumn{6}{c}{FS $N_{\rm WIN}$}\\
        \cmidrule{3-8}
        & & 128 & 256 & 512 & 1024 & 2048 & 4096 \\
        \midrule
        Coeff.  &  \SI{32}{\ms} & \SI{57}{\ms} & \SI{102}{\ms} & \SI{201}{\ms} & \SI{390}{\ms} & \SI{833}{\ms} & \SI{1795}{\ms} \\
        LP      &  \SI{29}{\ms} & \SI{58}{\ms} & \SI{98}{\ms} & \SI{195}{\ms} & \SI{376}{\ms} & \SI{804}{\ms} & \SI{1667}{\ms} \\
        LSTM 64 &  \SI{1322}{\ms} & - & - & - & - & - & - \\
        \bottomrule
    \end{tabular}
    }
    \label{tab:synth_bench}
    \vspace{-15pt}
\end{table}

Looking at the evaluation and benchmarking results, we observe that both configurations of our TD filter perform well and generally match or outperform the corresponding FS implementations.
Our method also provides roughly a 2x to 30x speedup over the FS and LSTM configurations.
The low-pass FS methods perform significantly worse when applied in the time domain which we attribute to overfitting of the neural network to the window size of the filter and can be clearly heard in the resulting audio.
This is less the case for the learned coefficient filters, especially according to the FAD metric. 
We hypothesise this could be due to the linear interpolation at inference of the learned coefficients which prevents harsh discontinuities from occurring at the frame boundaries, resulting in a smoother time-varying transfer function.
We also found during training that the Coeff FS synths would sometimes not converge, even when using gradient clipping, whereas our TD implementations never experienced stability issues.

\subsection{Modelling the LA-2A Leveling Amplifier}

For the compressor experiments, the targets we model are 1) the feed-forward compressor (FF) in Sec.~\ref{ssec:ff_comp} and 2) a Universal Audio LA-2A analog compressor (LA).
We optimise the proposed differentiable FF compressor ($\nabla$FF) to match the target sounds, examining its capability to replicate and infer the parameters of dynamic range controllers.
We test the following conditions:
\begin{itemize}
    \item FF-A: $R=3$, \SI{1}{\ms} attack and \SI{100}{\ms} release
    \item FF-B: $R=5$, \SI{30}{\ms} attack and \SI{30}{\ms} release
    \item FF-C: $R=8$, \SI{0.1}{\ms} attack and \SI{200}{\ms} release
    \item LA-D: compressor mode, 25 peak reduction
    \item LA-E: compressor mode, 50 peak reduction
    \item LA-F: compressor mode, 75 peak reduction
\end{itemize}

\begin{table}[h]
    \centering
    \caption{\itshape Summary of compressor ESR (\%) evaluation.}
    \vspace{-5pt}
    \resizebox{0.9\columnwidth}{!}{
        \begin{tabular}{ccccccc}
          \toprule
             Method &  FF-A&  FF-B&  FF-C&  LA-D& LA-E& LA-F\\
             \midrule
             FS&  2.362&  \bf 0.00780&  4.649& 11.29 & 9.485 & 7.783 \\
             $\nabla$FF& \bf 0.015&  0.00785& \bf 0.017& \bf 10.58 & \bf 9.356 & \bf 7.639\\
          \bottomrule
        \end{tabular}
    }
    \label{tab:comp_esr}
\end{table}

\begin{table}[h]
    \caption{\itshape The learned parameters for matching a LA-2A.}
    \vspace{-5pt}
    \label{tab:la2a_params}
    \centering
    \resizebox{1\columnwidth}{!}{
    \begin{tabular}{ccrrccrr}
        \toprule
          Method & Data &  $R$ & $CT$ (dB) & Attack & Release & $\alpha_{\rm rms}$ & $\gamma$ (dB) \\
          \midrule
          \multirow{3}{*}{FS}
          &D  &  9.1 & -11.66 &  \multicolumn{2}{c}{\SI{489.43}{\ms}}  &  0.008 & 0.69\\
          &E  &  231.1 & -19.08 &  \multicolumn{2}{c}{\SI{44.62}{\ms}}  &  0.606 & 0.34\\
          &F  &  2.9 & -26.00 & \multicolumn{2}{c}{\SI{0.06}{\ms}}   & 0.002 & -0.81\\
          \midrule
          \multirow{3}{*}{$\nabla$FF}
          &D  &  39.0 & -26.58 & \SI{99.41}{\ms} &  \SI{0.06}{\ms}  &  0.703 & 0.74\\
          &E  &  13.1 & -12.41 & \SI{5.68}{\ms}   &  \SI{420.56}{\ms} &  0.978 & 0.54\\
          &F  &  5.4 & -20.14 & \SI{2.24}{\ms}   &  \SI{229.15}{\ms} &  0.973 & -0.13\\
        \bottomrule
    \end{tabular}
    }
\end{table}

\begin{table}[h!]
    \centering
    \caption{\itshape Compressor runtime benchmarks for different lengths of \SI{44.1}{\kHz} audio. Ran on an M1 Pro MacBook for one optimisation step (forward + backward, one thread, single precision).}
    \vspace{-5pt}
    \resizebox{0.7\columnwidth}{!}{
    \begin{tabular}{cccc}
        \toprule
        \multirow{2}{*}{Method} & \multicolumn{3}{c}{Sequence duration}\\
         \cmidrule{2-4}
         &  \SI{30}{\second}&  \SI{60}{\second}& \SI{120}{\second}\\
         \midrule
         FS&  \SI{163.4}{\ms}&  \SI{320.8}{\ms}& \SI{663.8}{\ms}\\
         $\nabla$FF&  \SI{64.9}{\ms}&  \SI{117.9}{\ms}& \SI{239.4}{\ms}\\
         \bottomrule
    \end{tabular}
    }
    \label{tab:comp_bench}
\vspace{-5pt}
\end{table}

The $\alpha_{\rm rms}$, $CT$ and $\gamma$ for $\text{FF}_*$ are set to $0.03$, $-20$ and \SI{0}{\dB}, respectively. 
We train and evaluate our compressors on the SignalTrain dataset~\cite{hawley2019profiling}, which consists of paired data recorded in \\ \SI{44.1}{\kHz} from the LA-2A compressor with different peak reduction values.
Following~\cite{wright2022grey}, we select files with the sub-string \texttt{3c} in the file name.
Each LA-2A setting has \SI{20}{\minute} of paired audio containing real-world musical sounds and synthetic test signals.
We use the first \SI{5}{\minute} for training and the rest for evaluation.
The same input data is used for FF-A/B/C, and the target audio is generated by applying the FF compressor with the target parameters.
We pick the simplified compressor~\cite{steinmetz2022style} as our baseline, which uses the same FF compressor but has $\alpha_{\rm at} = \alpha_{\rm rt}$.
We denote this baseline as FS, as it uses frequency sampling to compute~\eqref{eq:rms} and~\eqref{eq:dyn_filter}.

The parameters we optimise are $\{\hat{R}, CT, \hat{\alpha}_{\rm at/bt/rms}, \gamma\}$.
We set $\alpha_* = \text{sigmoid}(\hat{\alpha}_*), R = \exp(\hat{R}) + 1$.
The initial values are \SI{50}{\ms} attack/release, $R=2$, $CT = \SI{-10}{\dB}$, $\alpha_{\rm rms} = 0.3$, and $\gamma = \SI{0}{\dB}$.
The conversion from time $t$ in seconds to coefficients $\alpha_*$ is $1 - \exp(-\frac{2.2}{44100t})$.
We train each compressor without using mini-batching for at least 1000 epochs using stochastic gradient descent with a learning rate of 100 and 0.9 momentum, minimising the mean absolute error (MAE).
For evaluation, we select the parameters with the lowest training loss.
We apply the same pre-filter from~\cite{wright2022grey} before calculating loss and evaluation metrics.
We use double precision when computing gigantic FFTs for the FS method to avoid numerical overflow.

Table~\ref{tab:comp_esr} shows that $\nabla$FF has a lower ESR than FS besides condition FF-B.
The parameters learned for condition FF-A/B/C using $\nabla$FF are close to the ground truth.
FS can only recover the parameters when attack and release are identical (FF-B).
For LA-2A, $\nabla$FF reasonably captures the analog characteristics (fast attack and slow release, shown in Table~\ref{tab:la2a_params}) of condition E/F.
We found FS tends to learn unrealistically large attack/release times and ratios, as seen in Table~\ref{tab:la2a_params}, likely due to its simplified design.

Table~\ref{tab:comp_bench} shows our method is two to three times faster than FS.
Training takes roughly \SI{43}{\minute} for FS and \SI{17}{\minute} for $\nabla$FF on an M1 Pro MacBook.

\section{Conclusion and Future Work}
In this work, we propose an efficient backpropagation algorithm
and implementation for an all-pole filter that can be used to model
time-varying analog audio systems end-to-end using gradient descent. 
We demonstrate its advantages over previous frequency sampling approximations by using it to model a phaser, a time-varying subtractive synthesiser, and a compressor.
Our method outperforms frequency sampling in accuracy and training efficiency, especially when using the systems at the sample-rate level.
We make our code and audio samples available and provide the trained audio effect and synth models in a VST plugin.

Our future work involves extending the backpropagation algorithm to work with differentiable initial conditions and applying it to relevant tasks.
We also plan on benchmarking the forward-mode differentiation of our filter, investigating its numerical stability, and extending our gradient derivation to higher-order optimisation use cases.

\vspace{-5pt}
\section{Acknowledgements}
Funded by UKRI and EPSRC as part of the ``UKRI CDT in Artificial Intelligence and Music'', under grant EP/S022694/1, and by the Scottish Graduate School of Arts \& Humanities (SGSAH). 

\begin{footnotesize}
\bibliographystyle{IEEEbib}
\bibliography{refs} 
\end{footnotesize}

\end{document}